\documentclass[%
twocolumn,
a4paper,
10pt,
superscriptaddress,
groupedaddress,
showpacs,preprintnumbers,
nobibnotes,
 amsmath,amssymb,
 aps,
prb,1
letterpaper,
floatfix,
]{revtex4-1}

\usepackage{url}
\usepackage{graphicx,rotating,amsmath,color,dsfont,natbib}

\usepackage[dvipdfm,bookmarks=false,colorlinks=true,breaklinks]{hyperref}
\usepackage[active]{srcltx}

\hypersetup{
    unicode=false,          
    pdftoolbar=false,        
    pdfmenubar=false,        
    pdffitwindow=false,     
    pdfstartview={FitH},    
    pdftitle={ChebyshevBdG},    
    pdfauthor={Lucian Covaci},     
    pdfnewwindow=true,      
    linkcolor=blue,          
    citecolor=blue,        
    filecolor=blue,      
    urlcolor=blue,           
    linkbordercolor={1 0 0}
}

\newcommand{\be}{\begin{equation}}
\newcommand{\ee}{\end{equation}}
\newcommand{\bea}{\begin{eqnarray}}
\newcommand{\eea}{\end{eqnarray}}
\newcommand{\up}{\uparrow}
\newcommand{\down}{\downarrow}
\newcommand{\bwt}{\begin{widetext}}
\newcommand{\ewt}{\end{widetext}}
\newcommand{\ham}{\mathcal{H}}

\newcommand{\ra}{\rangle}
\newcommand{\la}{\langle}
\newcommand{\bsb}{\begin{subarray}}
\newcommand{\esb}{\end{subarray}}
\newcommand{\largem}{\!\!}

\newcommand{\gb}{\bar{G}}
\newcommand{\tw}{\tilde{\omega}}

\newcommand{\vecv}[2]{
\left(\largem
 \begin{tabular}{c}
  $#1$ \\
  $#2$
  \end{tabular}
  \largem
\right)
}

\newcommand{\vech}[2]{
\left(\largem
 \begin{tabular}{c}
  $#1$  $#2$
  \end{tabular}
  \largem
\right)
}

\newcommand{\mat}[4]{
\left(
\largem
 \begin{tabular}{cc}
  $#1$ & $#2$ \\
  $#3$ & $#4$
  \end{tabular}
  \largem
\right)
}

\begin{document}
\title{Superconducting proximity effect in graphene under inhomogeneous strain}

\author{L. Covaci}
\affiliation{
Department Fysica, Universiteit Antwerpen,
Groenenborgerlaan 171, B-2020 Antwerpen, Belgium}
\author{F. M. Peeters}
\affiliation{
Department Fysica, Universiteit Antwerpen,
Groenenborgerlaan 171, B-2020 Antwerpen, Belgium}

\begin{abstract}
The interplay between quantum Hall states and Cooper pairs is usually hindered by the suppression of the superconducting state due to the strong magnetic fields needed to observe the quantum Hall effect. From this point of view graphene is special since it allows the creation of strong pseudo-magnetic fields due to strain. We show that in a Josephson junction made of strained graphene, Cooper pairs will diffuse into the strained region. The pair correlation function will be sub-lattice polarized due to the polarization of the local density of states in the zero pseudo-Landau level. We uncover two regimes; 1) one in which the cyclotron radius is larger than the junction length in which case the supercurrent will be enhanced, and 2) the long junction regime where the supercurrent is strongly suppressed because the junction becomes an insulator. In the latter case quantized Hall states form and Andreev scattering at the normal/superconducting interface will induce edge states. Our numerical calculation has become possible due to an extension of the Chebyshev Bogoliubov-de Gennes method to computations on video cards (GPUs).
\end{abstract}

\pacs{73.22.Pr,74.45.+c,73.43.-f}

\maketitle
Graphene has come in recent years to the forefront of condensed matter research not only due to its peculiar electronic properties but also due to its technological potential \cite{geim2009,castro_neto2009}. The special arrangement of carbon atoms in a honeycomb lattice has major consequences for its electronic properties. The particular way of coupling between lattice deformations and the electronic states results in remarkable properties. Any strain (either intrinsically due to phonons \cite{suzuura2002} or ripples \cite{guinea2008} or extrinsically due to applied stress \cite{levy2010,guinea2009,guinea2010,fogler2008,low2010,vozmediano2010} couples to the electronic degrees of freedom as a gauge field. The induced field does not break time reversal symmetry, because it has opposite sign in the K and K' Dirac cones. For this reason the field  is referred to as a pseudo-magnetic field. Electrons with momentum pertaining to different valleys will feel the effect of a reversed magnetic field by being deflected in opposite directions \cite{chaves2010}. More interestingly, it was recently shown \cite{guinea2009,low2010} that when the pseudo-magnetic field is slowly varying, the electronic spectrum shows the appearance of pseudo-Landau levels. Another special ingredient in our study is the existence of superconducting correlations in the graphene layer. While intrinsic superconductivity was not observed experimentally and only predicted theoretically in very specific conditions, e.g. the presence of high doping by adatoms \cite{uchoa2007}, superconducting correlations can be induced by proximity to a superconducting contact \cite{heersche2007,tomori2010,kanda2010,jeong2011}. Putting together these two effects one can achieve the coexistence of pseudo-quantum Hall states and superconducting correlations. Of special interest are the edge states formed at the interface between normal and superconducting regions since for energies below the superconducting gap quasiparticles will undergo Andreev scattering \cite{beenakker2008}. The coexistence between superconductivity and quantum Hall states was observed in superconductor/2DEG structures in magnetic fields. In order to observe quantized conductances the magnetic length has to be small when compared to the scattering length but still large when compared to the superconducting coherence length \cite{eroms2005,giazotto2005,hoppe2000}. The coexistence between antagonistic states can be achieved in graphene because the pseudo-magnetic field couples to the electronic orbital degrees of freedom in the graphene layer while leaving the superconducting contacts unaffected. One can envision the generation of edge states generated by large pseudo-magnetic fields (even on the order of hundreds of Tesla) but which undergo Andreev reflections at the superconducting/graphene interface. From this point of view, graphene is unique. In this paper we describe a strained graphene Josephson junction and show how the proximity effect can be tuned by inhomogeneously straining the junction and how the Josephson current could be carried either across the junction or by edge states.

\begin{center}
 \begin{figure}[t]
  \includegraphics[width=0.9\columnwidth]{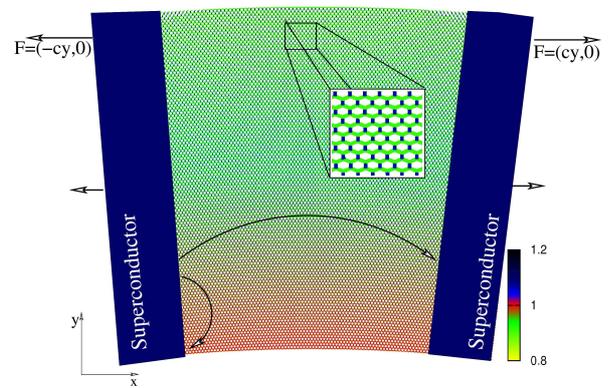}
  \caption{(Color online) Strained graphene Josephson junction obtained from a rectangle by applying linearly varying forces along the edges. The grayscale (color) code of the inter-atomic links is given by the ratio $t_{ij}/\gamma_0$ and shows that the links along $x$ are stretched while the ones along $y$ are compressed. Depending on the strength of the pseudo-magnetic field the cyclotron radius is such that it can take the quasiparticle across the junction or generate an edge state.} 
\label{fig1}
 \end{figure} 
\end{center}

We consider a Josephson junction made of strained graphene. Previous theoretical descriptions of graphene Joshepson junctions used both the low energy Dirac formulation\cite{titov2006} and tight-binding Hamiltonians \cite{black-schaffer2008}. Our model follows closely Ref. (\onlinecite{black-schaffer2008}) which consider that graphene under the contacts has an intrinsic pairing potential, then the Bogoliubov-de Gennes equations are solved self-consistently. To solve the problem we take instead another route by considering three dimensional superconducting contacts. Since the inverse proximity effect of the graphene layer on the contacts will be negligible and since there are no intrinsic superconducting correlations in graphene, a non self-consistent approach is reasonable. Our approach is also based on tight-binding Hamiltonians, therefore it will give similar results for the unstrained junction and not shine light in the conflicting results of Refs. (\onlinecite{titov2006}) and (\onlinecite{black-schaffer2008}). The system shown in Fig.~(\ref{fig1}) has contacts made of multilayer (10 layers) AA stacked graphite which is strongly coupled to the two-dimensional graphene sheet. Also from an experimental point of view, in order to have a sizable proximity effect the contact resistance between the contacts and the graphene layer has to be very low. For the purpose of our calculation, e.g. the superconducting proximity effect, the specifics of the contact band structure and coupling to the graphene sheet are not important. A detailed calculation which takes into account the coupling between a metallic gate and graphene could in principle be done but the quantitative description of the contacts will not add much in terms of qualitative description of the system. Such strained junctions could be experimentally achieved by rotating the superconducting contacts with respect to the symmetry axis of the junction.

The electronic properties are described by a tight-binding Hamiltonian for the $\pi$ carbon orbitals. The minimal Hamiltonian needed to describe superconducting correlations has the following Nambu spinor form:
\be
\label{eq:hamil1}
\ham=\sum_{\la i,j \ra} \vech{c_{i\up}^\dagger}{c_{i\down}} \hat{\ham}_{ij}
\vecv{c_{j\up}}{c_{j\down}^\dagger}
\ee
where $\hat{\ham}_{ij}$ is a $2 \times 2$ matrix:
\be
\label{eq:hamil2}
\hat{\ham}_{ij}=\mat{-\mu}{\Delta_i}{\Delta_i^\star}{\mu}
\delta_ { ij } +\mat {-t_{ij}} {0}
{0}{t_{ij}^\star}(1-\delta_{ij}).
\ee
where the sum $\la i,j \ra$ is over nearest and next nearest neighbors. The superconducting order parameter, $\Delta_i$, is of s-wave spin-singlet type and is non-zero only in the contact region above the graphene sheet. The strain information is included in the modified hopping amplitudes $t_{ij}$ according to the empirical relation $t_{ij}=\gamma_0 \exp^{3.37(\frac{r_{ij}}{a_0}-1)}$, where $\gamma_0=3eV$ and $a_0$ is the equilibrium inter-carbon distance \cite{pereira2009}. This also gives a good approximation for the next-nearest neighbor hopping amplitude. In order to describe the inhomogeneous strain we consider the deformed rectangular geometry shown in Fig.~(\ref{fig1}) which can be obtained from a rectangle by applying a linearly varying force $\vec{F}_L= (-const*y,0)$ on the left side and $\vec{F}_R=(const*y,0)$ on the right side \cite{guinea2010}. We obtain the strain tensor by numerically solving the elasticity problem for an isotropic elastic sheet with the elastic properties of an ideal flat graphene sheet.

Since we consider three dimensional contacts and the strain is inhomogeneous it is not possible to use any symmetries in the calculations of the electronic properties. Therefore the system size needed to describe this problem becomes prohibitively large for an exact diagonalization of the Hamiltonian. Instead we will numerically obtain an approximation of the Gorkov Green's function by using the Chebyshev Bogoliubov de Gennes (CBdG) method \cite{covaci2010}. The $2\times2$ Green's function is defined as:
\be
\label{eq:green1}
\bar{G}_{ij}(\omega)=\la vac | \vecv{c_{i\up}}{c_{i\down}^\dagger}
\hat{G}(\omega) \vech{c_{j\up}^\dagger}{c_{j\down}} |vac \ra 
\ee
where $\hat{G}(\omega+i \eta)=[\omega + i\eta-\ham]^{-1}$, $ |vac \ra$ is the vacuum and the diagonal(normal) and off-diagonal(anomalous) components can be expressed as:
\bea
\label{eq:green2}
\bar{G}_{ij}^{11}(\omega)&=&\la c_{i\up}|\hat{G}(\omega)|c_{j\up}
^\dagger \ra \\
\bar{G}_{ij}^{12}(\omega)&=&\la
c_{i\down}^\dagger|\hat{G}(\omega)|c_{j\up }^\dagger \ra^\ast.
\eea
First a scaling of the excitation energies is performed, e.g. $\tilde{\ham}=(\ham-\mathds{1}b)/a$, $\tilde{\omega}=(\omega-b)/a$ where $a=(E_{max}-E_{min})/(2-\eta)$ and $b=(E_{max}+E_{min})/2$, where $\eta>0$ is a small number. Following Refs.~[\onlinecite{covaci2010,weise2006}], the Green's function's components can be expressed as an expansion written in terms of Chebyshev polynomials:
\be
\label{eq:gf1}
\gb_{ij}^{11(12)}(\tw)=\frac{-2 i}{\sqrt{1-\tw^2}} \sum_{n=0}^\infty a_n^{11(12)}(i,j)
e^{-i\: n \cdot \arccos(\tw)}
\ee
where the coefficients $a_n^{11(12)}(i,j)$ can be obtained by an iterative procedure involving repeated applications of the Hamiltonian on iterative vectors. The physical properties that can be straightforwardly extracted from the Green's functions are the local density of states (LDOS) $N_i(\omega)=-\frac{1}{\pi} Im[\gb_{ii}^{11}(\omega)]$ and the spin-singlet superconducting pair correlation function $<c_{i\up} c_{i\down}> = i \int_{-E_c}^{E_c} \gb^{12}_{ii}(E) (1-2f(E)) dE$.

Even though it is feasible to use the CBdG method on parallel computer clusters even for a large number of atoms \cite{covaci2010} (even up to one million) we have instead developed codes which take advantage of the parallel nature of Graphics Processing Units(GPU). Significant speedups on the order of 1000 can be achieved for calculating electronic properties of systems described by tight-binding Hamiltonians; the system size is on the order of hundreds of thousands of atoms. The calculations presented here have been performed on a system containing three Nvidia Geforce GTX580 which ran in parallel.

\begin{center}
 \begin{figure}[ttt]
  \includegraphics[angle=-90,width=\columnwidth]{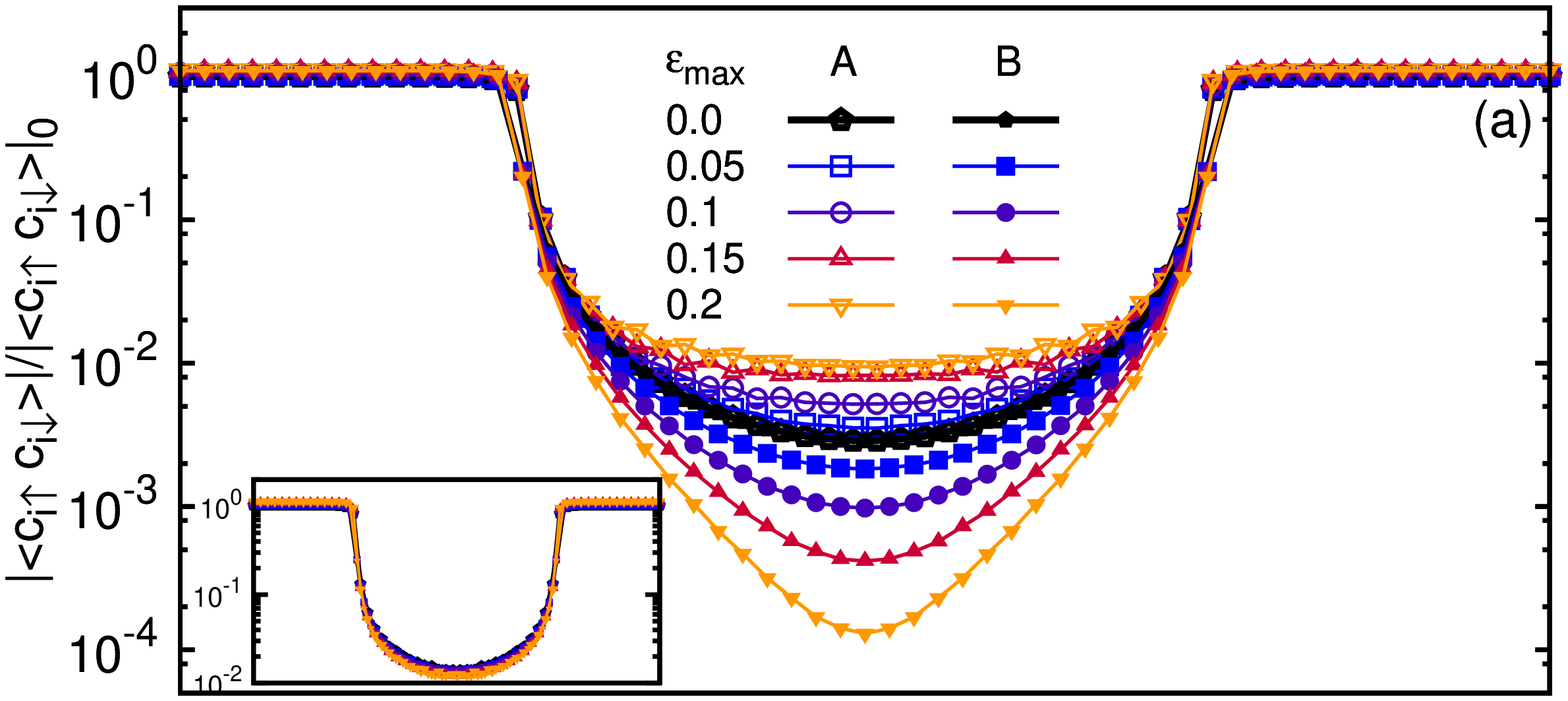}
  \includegraphics[angle=-90,width=\columnwidth]{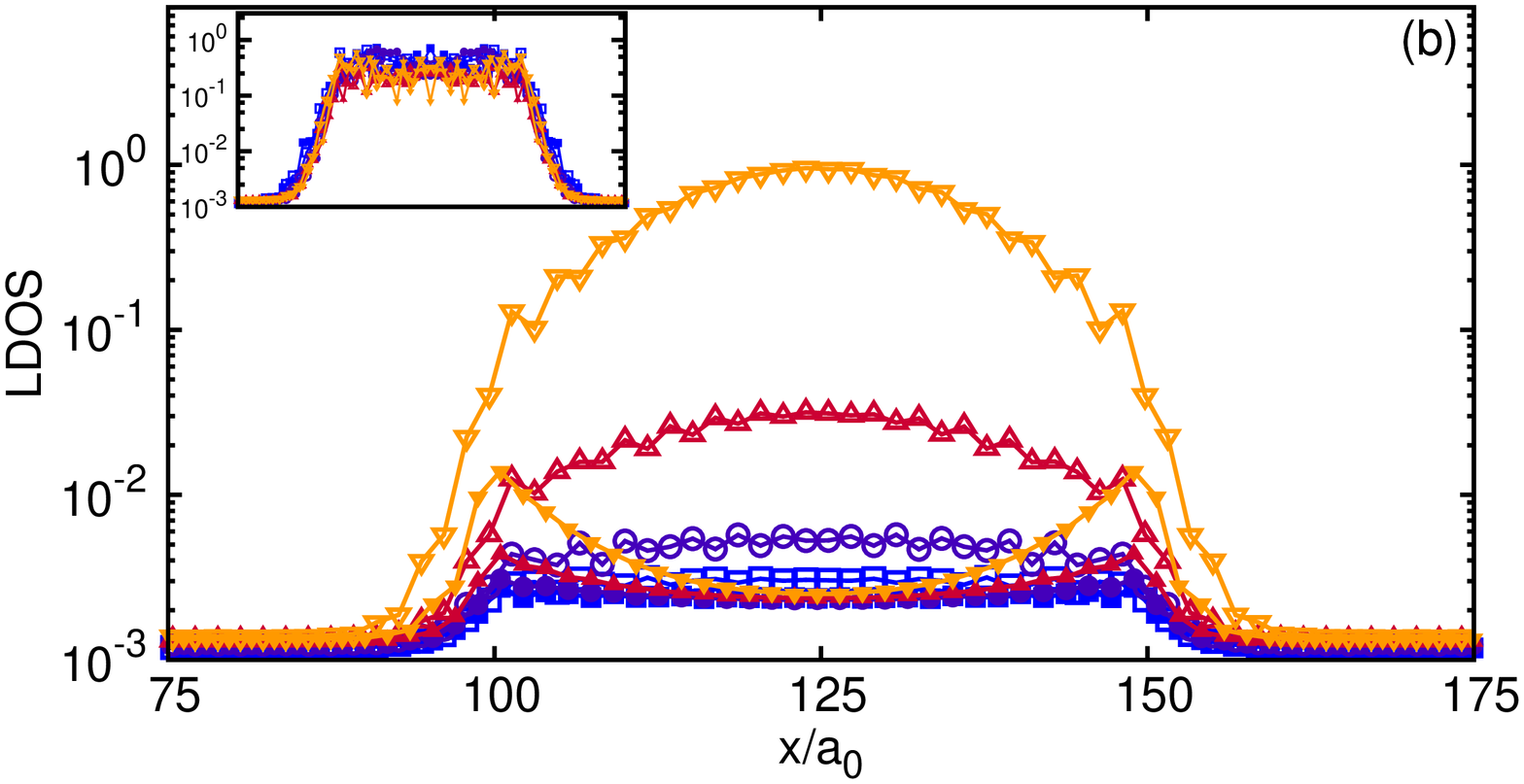}
  \caption{(Color online) a) Absolute value of the pair correlation function along a line, $y=L_y/2$, as a function of position for various strain configurations for $\mu=0.24\gamma_0$ (the zero pseudo-Landau level is at the Fermi level). The inset shows the pair correlation function for $\mu=0$. b) Local density of states along the same line as a function of position for $\mu=0.24\gamma_0$. The inset shows the LDOS for $\mu=0$.} 
\label{fig2}
 \end{figure} 
\end{center}

In Fig.~(\ref{fig2}a) we show the pair correlation function as a function of position for various strain configurations for $\mu=0.24\gamma_0$. The plot is taken along a line $y=L_y/2$. The maximum strain, $max(u_{xx})=\epsilon_{max}$, gives the curvature of the deformed junction and is a direct measure of the pseudo-magnetic field \cite{guinea2010,low2010}. As expected, under the superconducting contact the pair correlation is constant while in the junction it decreases exponentially. Notice that as the strain increases the results for the two sub-lattices start to differ. While for one sub-lattice the pair correlation is enhanced when compared to the unstrained junction, in the other sub-lattice it is strongly suppressed. The chemical potential is chosen such that the zero pseudo-Landau level is pinned at the Fermi level (since we also consider next nearest neighbors it is shifted away from $\mu=0$). The inset shows the same quantity but for $\mu=0$ and shows no significant difference on the two sub-lattices. In order to understand this peculiar behavior we also plot in Fig.~(\ref{fig2}b) the LDOS for the same configurations. We see a breaking of the sub-lattice symmetry which shows that the LDOS is large in one sub-lattice and suppressed in the other. The symmetry breaking of the zero-th pseudo-Landau level can be observed even in a strained graphene sheet with no superconducting contacts and is a direct consequence of the opening of a gap. It can be induced by breaking either the time reversal symmetry or the sub-lattice symmetry and since the pseudo-magnetic fields have no net flux (they have alternating sign in the two Dirac cones) the opening of the gap can be achieved by breaking the symmetry of the zero pseudo-Landau level. Changes in the super-current flow in strained Josephson junctions are expected; flow will only occur in one sub-lattice (A). The inset of Fig.~(\ref{fig2}b) shows the LDOS for $\mu=0$ and since the Fermi level is away from the zero pseudo-Landau level there is no symmetry breaking and the proximity effect is a conventional one.

\begin{center}
 \begin{figure}[ttt]
  \includegraphics[angle=-90,width=\columnwidth]{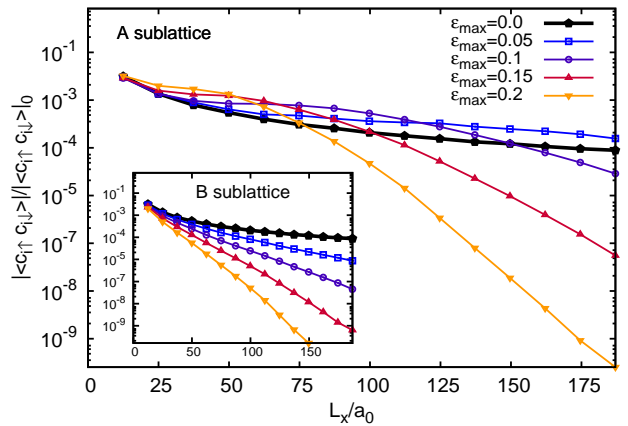}
  \caption{(Color online) Pair correlation at $x=L_x/2$, $y=L_y/2$ as a function of $L_x$ for different strain configurations. For the A sub-lattice two regimes are found depending on the ratio $r_c/L_x$. Inset: the same but now for the B sub-lattice where the pair correlation decays exponentially with $L_x$.} 
  \label{fig3}
  \end{figure} 
\end{center}

The dependence of the pair correlation at the mid point $(L_x/2,L_y/2)$ for $\mu=0.24\gamma_0$ as a function of the length of the junction is shown in Fig.~(\ref{fig3}) where two regimes are found. First, when the junction is shorter than twice the effective cyclotron radius, $r_c \propto \sqrt{1/\epsilon_{max}}$, quasi-particles with energy lower than the superconducting gap are bound and after undergoing Andreev scattering at the normal/superconducting interface they will scatter between the contacts. In this regime the pair correlation in the A sub-lattice is enhanced, because the LDOS is itself enhanced by the pseudo-quantum Hall effect. The diffusion length of Cooper pairs is larger therefore enhancing the pair correlations at $L_x/2$. Second, when the junction is longer than twice the effective cyclotron radius quantized states will form in the junction which becomes insulating. Andreev scattering will only affect edge states near the normal/superconducting interface. In this regime the diffusion length is strongly suppressed and the pair correlation at $L_x/2$ shows an exponential decay. The inset of Fig.~(\ref{fig3}) shows that in the B sub-lattice the diffusion length of the Cooper pairs decreases monotonically with the strain and the pair correlation at $L_x/2$ decays exponentially with $L_x$ faster than the pair correlation in the A sub-lattice due to the absence of states localized in the B sub-lattice.

\begin{center}
 \begin{figure}[ttt]
  \includegraphics[angle=-90,width=\columnwidth]{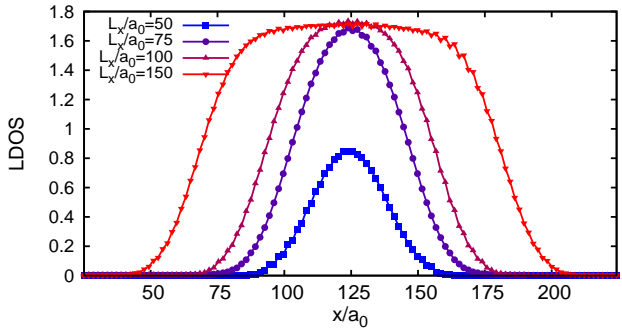}
  \caption{(Color online) LDOS at the Fermi level for the A sub-lattice across the junction at $y=L_y/2$ for different $L_x$.} 
\label{fig4}
 \end{figure} 
\end{center}

These two regimes are also present when considering the formation of quantum Hall states in the presence of pseudo-magnetic fields. This is shown in Fig.~(\ref{fig4}) where we plot the LDOS at the Fermi level for the A sub-lattice for the same strain configuration, $\epsilon_{max}=0.2$, but for different lengths of the junction. For short junctions the quasi-particles scatter from both contacts and the formation of quantized orbits is hindered. For long junctions we observe a plateau in the LDOS signaling the appearance of quantized states in the center of the junctions. In this case the Andreev scattering affects only a narrow region near the normal/superconducting interface.

\begin{center}
 \begin{figure}[ttt]
  \includegraphics[angle=0,width=\columnwidth]{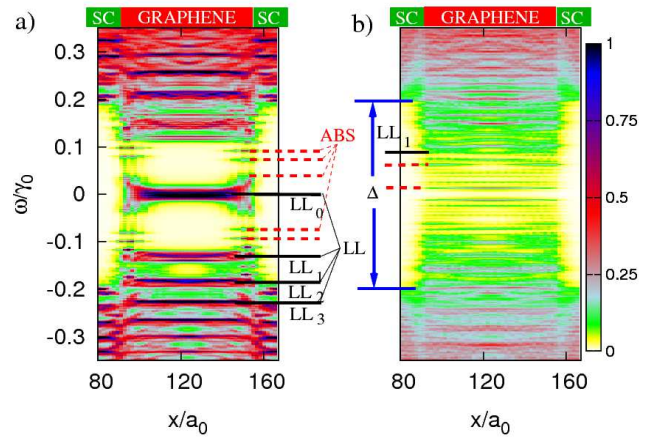}
  \caption{(Color online) LDOS across the junction at $y=L_y/2$ for a) $\epsilon_{max}=0.3$ and b) $\epsilon_{max}=0.1$. The value of the order parameter in the contacts is $\Delta=0.2\gamma_0$ and the chemical potential is $\mu=0.24\gamma_0$. As a guide to the eye we show the position of the LL(ABS) by horyzontal black(red) lines and the superconducting gap ($\Delta$).} 
\label{fig5}
 \end{figure} 
\end{center}

To further see how the pseudo-Landau levels disperse near the interface we plot in Fig.~(\ref{fig5}) the LDOS across the junction for fixed $y=L_y/2$ and atoms in the A sub-lattice for three strain configurations. When the pseudo-magnetic field is strong ($\epsilon_{max}=0.3$) we observe the formation of pseudo-Landau levels (LL) (up to $n=5$) which are localized in the junction. They disperse only very close to the normal/superconducting junctions ($L_x/a_0=75$ and $L_x/a_0=175$) and even show quantized edge states due to Andreev reflections, i.e. these are Andreev Bound States (ABS). In the B sub-lattice the LDOS is similar except at the Fermi level where the zero pseudo-Landau level is missing. As the strain is reduced to $\epsilon_{max}=0.2$ and $\epsilon_{max}=0.15$ the pseudo-Landau levels become dispersive away from the interfaces and in the latter case quantized states due to Andreev reflections can be observed throughout the whole junction.

In conclusion we have shown that by using an efficient numerical method (CBdG) which is extended to run on GPUs with the benefit of large speedups, simulations of strained graphene junctions with a large number or atoms are possible. We showed for the first time that strained graphene proves to be a playground for testing the exotic coexistence between strong pseudo-magnetic fields and superconducting correlations. This is due to the fact that the pseudo-magnetic fields only couple to the orbital motion of electrons in the graphene sheet and not to their spin. Any superconducting contact will be unaffected by these strong pseudo-magnetic fields. 
By considering a Josephson junction made of strained graphene we showed that due to the proximity effect Cooper pairs can leak into the strained region. In addition, due to the peculiar nature of the pseudo-Landau levels the pair correlation function has a broken sub-lattice symmetry, being enhanced in one sub-lattice and suppressed in the other. We have also uncovered two regimes depending on the ratio between the cyclotron radius and the junction length; in one regime quasi-particles will bounce between the superconducting contacts in which case the Josephson current would be enhanced while in the other regime quasi-particles scatter only from one contact thus creating an Andreev bound edge state in which case the Josephson current would be strongly suppressed across the junction. In this case the supercurrent can only propagate through edge states. From an experimental point of view, the particular strain configuration considered here would not be a necessary ingredient. Instead what is of utmost importance is the existence of a finite pseudo-magnetic field which could be achieved by any inhomogeneous strain configuration which generates the pseudo-Landau levels.

Acknowledgments:  This work was supported by the Flemish Science Foundation (FWO-Vl) and the Euro GRAPHENE project CONGRAN. Discussions with Andrey Chaves are gratefully acknowledged. 

%

\end{document}